\pdfoutput=1

\documentclass[
  twocolumn,
  prb,
  showpacs,
  amsmath,
  amssymb,
  superscriptaddress
]{revtex4}

\usepackage{bm}
\usepackage{graphicx}
\usepackage{amssymb}
\usepackage{color}
\usepackage{braket}
\usepackage{empheq}
\newcommand{\tr}{\text{tr}}
\renewcommand{\v}[1]{\textbf{\textit #1}}

\begin{document}

\title{Kerr effect from diffractive skew scattering in chiral $p_x \pm i p_y$ superconductors}

\author{Elio J.\ K\"onig and Alex  Levchenko \\ \textit{Department of Physics, University of Wisconsin-Madison, Madison, Wisconsin 53706, USA}}

\begin{abstract}
We calculate the temperature dependent anomalous ac Hall conductance $\sigma_H(\Omega, T)$ for a two-dimensional chiral $p$-wave superconductor. This quantity determines the polar Kerr effect, as it was observed in Sr$_2$RuO$_4$ [J. Xia \textit{et al.}, Phys.~Rev.~Lett.~\textbf{97}, 167002 (2006)]. We concentrate on a single band model with arbitrary isotropic dispersion relation subjected to rare, weak impurities treated in the Born approximation. As we explicitly show by detailed computation, previously omitted contributions to extrinsic part of an anomalous Hall response, physically originating from diffractive skew scattering on quantum impurity complexes, appear to the leading order in impurity concentration. By direct comparison with published results from the literature we demonstrate the relevance of our findings for the interpretation of the Kerr effect measurements in superconductors.
\end{abstract}

\date{\today}


\maketitle

\textbf{\textit{Introduction.}}
Unconventional superconductivity remains a very active field of condensed matter research. Notably, the chiral $p$-wave superconductor is a particularly spectacular state of matter. Not only it demonstrates the extraordinary effects of electronic correlations, but it also displays exciting topological features, such as Majorana zero modes bound to half quantum vortices. 
In a chiral $p$-wave superconductor, the electrons which constitute the Cooper pairs rotate around each other with magnetic quantum number $L_z = \pm 1$. Clearly, such a state breaks time-reversal symmetry (TRS) and by Pauli's exclusion principle, the Cooper pair wave function ought to be symmetric in spin or band indices of a given material. 

To present date, the chiral $p$-wave superconducting phase has not yet been unambiguously observed experimentally in solids. Nonetheless, there is wide consensus in the community, that strontium ruthenate (Sr$_2$RuO$_4$) constitutes a promising candidate material.\cite{MaenoLichtenberg1994,MaenoMacKinzie2003,Mineev2010,MaenoIshida2011,LiuMao2015,KallinBerlinsky2016} Experimental evidence for triplet-pairing in Sr$_2$RuO$_4$ relies on the Knight shift\cite{IshidaMaeno1998} and neutron scattering\cite{DuffyMcIntyre2000} while a peculiar phase sensitivity of the Josephson effect \cite{NelsonMaeno2004} is believed to reveal the odd parity of the order parameter. Furthermore, the observation of half quantum vortices in magnetometry\cite{JangMaeno2011} indicates spin triplet $p$-wave superconductivity. The spontaneous breaking of TRS was first observed in the muon spin-relaxation\cite{LukeSigrist1998} and later in the polar Kerr effect (PKE).\cite{XiaKapitulnik2006} In this paper, we concentrate on the latter probe.
A nonzero Kerr angle 
\begin{equation}
\theta_K = \frac{4\pi}{\Omega d} \Im \left [\frac{\sigma_{H}(\Omega)}{n(n^2 - 1)}\right ]
\end{equation}
in a layered material (such as Sr$_2$RuO$_4$) with interlayer distance $d$ and complex index of refraction $n$ relies on a finite, 2D, optical anomalous Hall conductivity $\sigma_H(\Omega) = [\sigma_{xy} (\Omega) -\sigma_{yx} (\Omega)]/2$, with $\Omega$ the ac frequency.
%

\begin{figure}[t]
\begin{center}
\includegraphics[scale=.5]{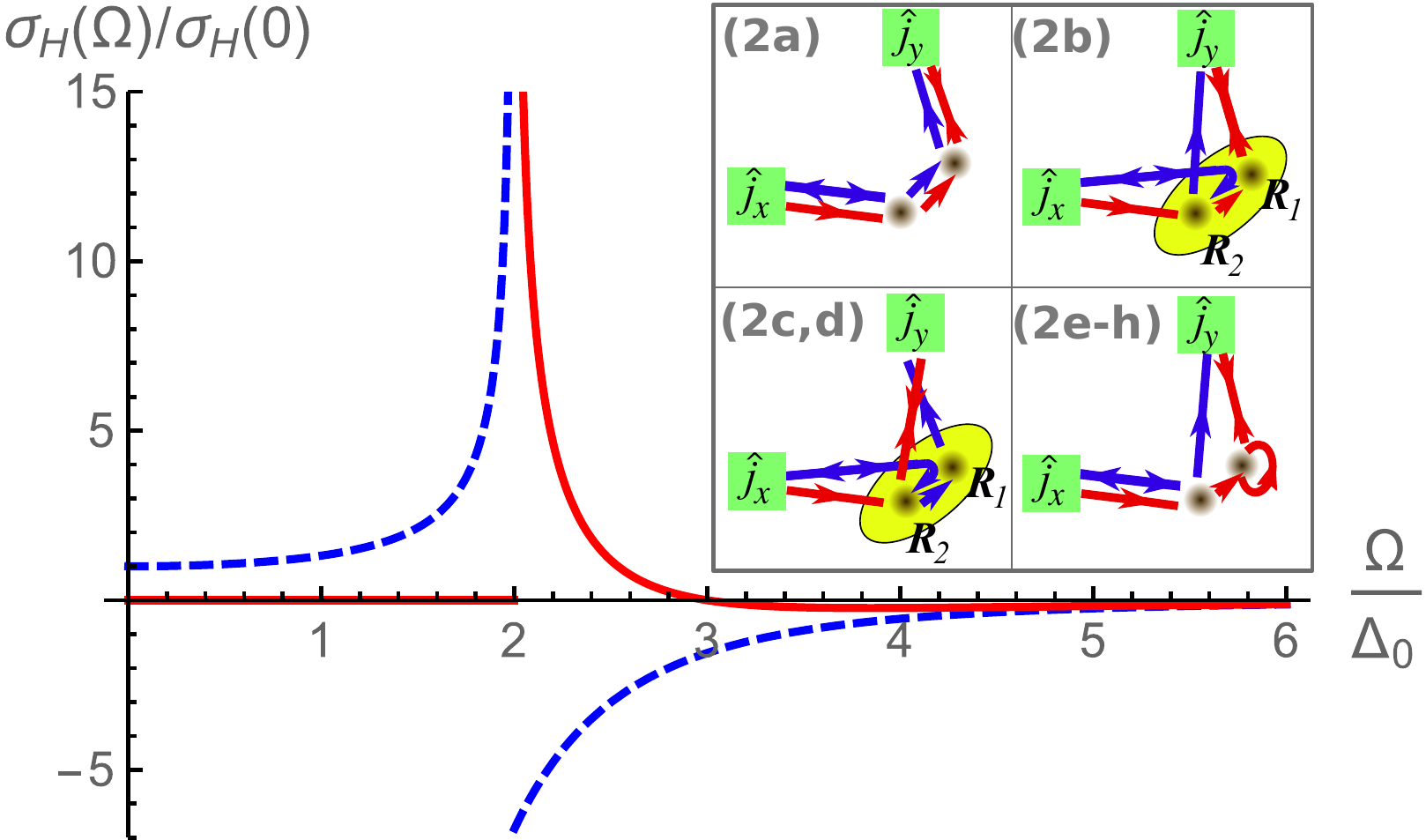} 
\caption{Zero temperature Hall conductivity for a chiral $p_x \pm i p_y$ superconductor with weak impurities and, for concreteness, a quadratic dispersion relation (ac frequency $\Omega$,  superconducting pairing amplitude $\Delta_0$, elastic scattering time $\tau$ and $\sigma_H(0) = \mp e^2/[105 \pi (\Delta_0 \tau)^2 \hbar]$). The solid red (blue dashed) curve represents the imaginary (real) part of $\sigma_H$. Inset:  Real space illustration of quantum mechanical probabilities for processes contributing to $\sigma_H$ and corresponding to diagrams (2a)-(2h) from Fig.~\ref{fig:PertDiagsSecond}. While generally all of those diagrams contribute, in the specific case of a parabolic band, the response stems from the processes (2b)-(2d), only. These contributions rely on diffractive scattering from quantum impurity complexes (yellow ellipses) with spatial extension  $\vert \v R_1 - \v R_2 \vert$ comparable to the Fermi wavelength $\lambda_F$. }
\label{fig:HallPlot}
\end{center}
\end{figure}

Theories of the anomalous Hall effect\cite{NagaosaOng2010} (AHE)  are most often developed on the basis of either the semiclassical Boltzmann equation\cite{Sinitsyn2007} or the Kubo-Streda\cite{Streda1982} diagrammatic formalism. While both approaches are equally justified and should yield the same results,\cite{SinitsynSinova2007} the semiclassical approach seems to be more intuitive while diagrams appear to be more systematic. In the Boltzmann treatment, the AHE is attributed to the addition of the following effects. First, the intrinsic or anomalous velocity contribution, which relies on the Berry curvature of the bands. Second, the extrinsic contributions, which stem from (a) asymmetric skew scattering from impurities, and (b) the side jump, a lateral displacement of semiclassical trajectories near scattering centers. These contributions are automatically accounted for in the diagrammatic treatment of the problem. Most recently, the importance of diagrams with two crossed impurity lines\cite{SinitsynSinova2007} was uncovered. \cite{AdoTitov2015,AdoTitov2016} Physically, these diagrams represent diffractive skew scattering from quantum impurity complexes.\cite{KoenigLevchenko2016} It is important to emphasize, that for a disorder potential with Gaussian distribution, diagrams with two crossed impurity lines are of the same order as diagrams within the noncrossing approximation.

\begin{figure}
\begin{center}
\includegraphics[scale=.3]{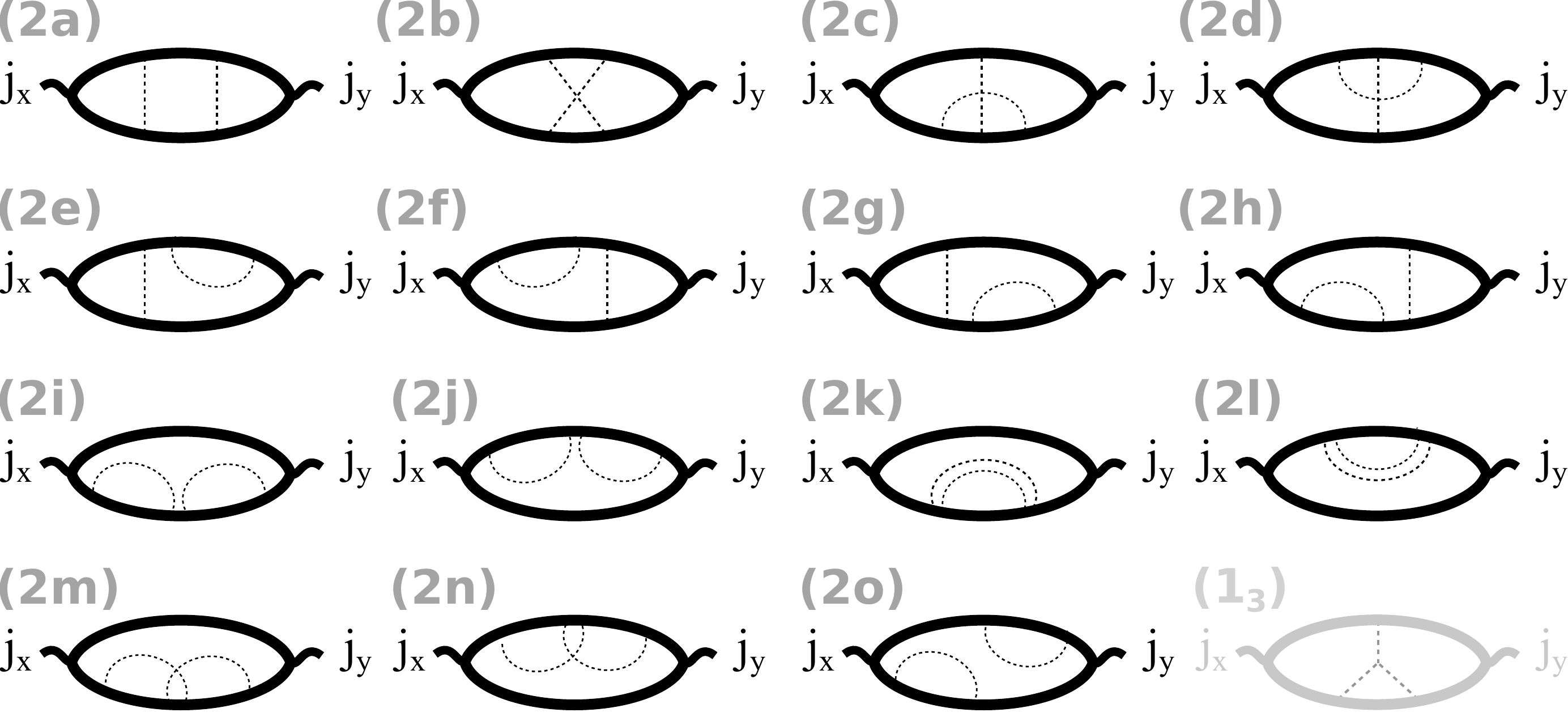} 
\caption{Diagrams (2a)-(2o): $\sigma_{H}(\Omega)$ to second order in impurity concentration for the model defined by Eqs.~\eqref{eq:Model}. Diagrams (2a) and (2e)-(2h) were presented in Ref.~[\onlinecite{LutchynYakovenko2009}]. Diagrams (2i)-(2o) are zero. Diagrams (2b)-(2d) are the diffractive contributions which are the major focus in this work. Diagram $(1_3)$: ``Mercedes star" diagram\cite{Goryo2008,LutchynYakovenko2009} occuring for a model with non-Gaussian disorder.}
\label{fig:PertDiagsSecond}
\end{center}
\end{figure}

Theoretically, the ac AHE in the context of chiral $p$-wave superconductors has been studied in Refs.~\onlinecite{Volovik1988b,Volovik1988,Yakovenko2007,Mineev2007,Mineev2008,RoyKallin2008} for clean single band models. However, $\sigma_H(\Omega) = 0$ for such models,\cite{RoyKallin2008,LutchynYakovenko2009} a result that can be understood as a consequence of Galilean invariance.\cite{ReadGreen2000} Therefore, the observed finite Kerr effect was considered within clean multiband models\cite{WysokinskiGyorffy2012,Mineev2012,TaylorKallin2012,TaylorKallin2013,GradhandGyorffy2013} and single band models with impurities.\cite{Goryo2008,LutchynYakovenko2009,LiAndreevSpivak2015} Notwithstanding the significant theoretical interested, to the best of our knowledge the effect of diffractive skew scattering from quantum impurity complexes has been disregarded in the literature, so far. It will therefore be the subject of the present paper. We concentrate on a single band model for a chiral $p$-wave superconductor and treat weak impurities perturbatively and in the Gaussian (i.e. Born) approximation. In this case, the contribution to the zeroth and first order in the impurity concentration vanishes. We will show that diffractive skew scattering, represented by crossed diagrams (2b-2d) of Fig.~\ref{fig:PertDiagsSecond}, contributes to the same order as diagrams in the noncrossing approximation, (2a) and (2e-2o) in Fig.~\ref{fig:PertDiagsSecond}.

\textbf{\textit{Model and Assumptions.}}
We employ the following 2D mean-field Bogoliubov-de Gennes Hamiltonian
\begin{subequations}
\begin{equation}
H_0 = \xi_{\v p} \tau_z  + \frac{\Delta_0}{p_F} (p_x \tau_x + \zeta p_y \tau_y) \label{eq:H0}
\end{equation}
to describe the single band chiral $p$-wave superconductor under consideration. Here, $\Delta_0$ is the mean-field superconducting amplitude, $p_F$ is the Fermi momentum and $\zeta= \pm 1$ determines the chirality of the superconductor. Pauli matrices in Nambu space are denoted by $\tau_{x,y,z}$. The dispersion relation (DR) $\xi_{\v p}$ is assumed to be isotropic $\xi_{\v p} = \xi_p$. While we derive and present all results for a generic DR, we will additionally discuss our findings for a parabolic band $\xi_{\v p} = p^2/2m - \mathcal E_F$.  We remind the reader, that Sr$_2$RuO$_4$ is a layered material and that the conduction mainly takes place in the Ru-O planes. The model Hamiltonian~\eqref{eq:H0} should be a good description of the cylindrical $\gamma$-sheet in Sr$_2$RuO$_4$. \cite{MaenoMacKinzie2003}

In addition to Eq.~\eqref{eq:H0} our model contains point-like impurities of strength $u_0$ and density $n_{\rm imp}$ that we treat in the Born approximation. Then, the disorder potential $V(\v r)$, which enters the Hamiltonian as
\begin{equation}
H_{\rm dis} = V(\v r) \tau_z ,
\end{equation}
follows to have a Gaussian white noise distribution 
\begin{equation}
\langle V(\v r ) V(\v r') \rangle = \frac{\delta(\v r - \v r')}{2\pi \nu_0 \tau} = n_{\rm imp} u_0^2 \delta(\v r - \v r').
\end{equation}
\label{eq:Model}
\end{subequations}
In our notation, $\nu_0$ is the density of states (DOS) at the Fermi level and $\tau$ the elastic  scattering time, both taken in the normal phase. 

We consider a superconductor in the BCS limit in a degenerate electron gas with rare impurities. These assumptions correspond to the following hierarchy of energy scales:
\begin{equation}
\frac{1}{\tau} \ll \{\Delta_0, T, \Omega\} \ll \frac{v_F p_F}{2} \equiv \mathcal E_F.
\end{equation}
Here, $T$ is the temperature, $\Omega$ the ac frequency, $v_F$ the Fermi velocity and we set Boltzmann's and Planck's constants as well as the speed of light to unity $k_B = \hbar = c =1$. Our calculations are perturbative in impurity concentration, with the leading contributions being of second order. Furthermore, we keep only terms up to zeroth order in the small parameter $\alpha = {[\max ( \Delta_0, T, \Omega)]}/{\mathcal E_F} \ll 1$.

\textbf{\textit{Calculation.}}
Since all diagrams to zeroth and first order in impurity concentration vanish,\cite{LutchynYakovenko2009} we concentrate on second order contributions, see Fig.~\ref{fig:PertDiagsSecond}. 

We are interested in the response of the $p$-wave superconductor to a vector potential, which slowly varies on the length scale of the coherence length. Such a slow vector potential does not enter the momentum dependent order parameter. \cite{Volovik1988,LutchynYakovenko2009} The physical reason is that slow electromagnetic fields can not resolve the relative momentum of the electrons forming the Cooper pair. Technically, this is a consequence of $\mathbf U(1)$ gauge invariance, keeping in mind that the order parameter field transforms as a bilinear of two creation operators.
The current vertex is thus (electron charge $e = - \vert e \vert$)
\begin{equation}
\hat j_\mu =  e \v v_\mu(\v p) \mathbf 1_\tau =  e \frac{1}{2\pi \nu_0} \v p_\mu \mathbf 1_\tau .
\end{equation}
For a generic DR, the last equation is valid to leading order in $\alpha$ while it is exact for a parabolic band.

We now outline the calculation of the ac Hall response, more details can be found in Ref.~[\onlinecite{SuppMat}]. We use the Matsubara Green's functions
\begin{subequations}
\begin{eqnarray}
G(\epsilon_n, \v p) &=& [i \epsilon_n - H_0(\v p)]^{-1} = N_\v p(\epsilon_n) \mathcal G(\epsilon_n, \v p),
\end{eqnarray}
with fermionic frequency and momentum  $(\epsilon_n, \v p)$ and
\begin{eqnarray}
&&\hskip-.5cm
N_{\v p} (\epsilon_n) = -\left[i \epsilon_n + \xi_\v p \tau_z + \frac{\Delta_0}{p_F} (p_x \tau_x + \zeta p_y \tau_y) \right], \\
&&\hskip-.5cm
\mathcal G(\epsilon_n, \v p) = \frac{1}{\epsilon_n^2 + \xi_{\v p}^2 +  (\v p\Delta_0 /p_F)^2 }.
\end{eqnarray}
\label{eq:cleanGF}
\end{subequations}
We also need real space expressions for the Green's function and for $(\xi \mathcal G)(\epsilon_n, \v r) = \xi_{(-i \nabla)} \; \mathcal G(\epsilon_n, \v r)$ to order $\mathcal O(\alpha^0)$
\begin{subequations}
\begin{eqnarray}
&&\mathcal G(\epsilon_n,\v r)= \frac{ \pi\nu_0}{\sqrt{\epsilon_n^2 + \Delta^2}}  \left \lbrace J_0(p_F r) + \mathcal O (\alpha)\right \rbrace,\label{eq:realspaceGF} \\
&&(\xi \mathcal G)(\epsilon_n,\v r) = - {\pi \nu_0} \left  \lbrace F(p_F r) + \mathcal O(\alpha) \right \rbrace . \label{eq:realspacexiGF}
\end{eqnarray}
\end{subequations}
In Eq.~\eqref{eq:realspaceGF}, $J_0(p_F r)$ denotes the zeroth Bessel function of the first kind. The dimensionless function $F(p_F r)$ represents an off-shell contribution and therefore depends on microscopic details of the model. In the case of a quadratic dispersion we find $F(p_F r) = Y_0(p_F r)$, where $Y_0(p_F r)$ is the zeroth Bessel function of the second kind. These expressions are valid for length scales $r \ll v_F/\Delta_0$, i.e. the regime of length scales which of relevance for nonvanishing diagrams (2a)-(2h), see e.g. Eq.~\eqref{eq:beta}.

The transverse current-current correlator $Q_H (\omega_l)$ is evaluated at finite photon frequency $\omega_l$. We first evaluate diagrams (2i-2o) of Fig.~\ref{fig:PertDiagsSecond}. In view of the antisymmetrization $\sigma_{H}(\Omega) = [\sigma_{xy}(\Omega) - \sigma_{yx}(\Omega)]/2$ it is readily seen that diagrams (2i-2o) identically vanish after angular momentum integration. 

We next concentrate on the other diagrams in the noncrossing approximation. Diagram (2a) contributes
\begin{eqnarray}
Q_H^{(2a)}(\omega_l)  &=& \zeta \frac{e^2 \Delta_0}{2 \omega_l^2 (2 \tau)^2} \beta_{\rm FS} k(\omega_l). \label{eq:Q2amain}
\end{eqnarray}
We expanded the density of states near the Fermi surface as $\nu(\xi) \simeq \nu_0 + \nu_0' \xi$ and introduced the dimensionless constant
\begin{equation}
\beta_{\rm FS} = \frac{\mathcal E_F \nu_0'}{\nu_0}
\end{equation}
as well as the function
\begin{align}
 k(\omega_l) &= \sum_n \Bigg \lbrace \frac{T \Delta_0}{2 \epsilon_n + \omega_l}  \left [\sqrt{(\epsilon_n + \omega_l)^2 + \Delta_0^2} - \sqrt{\epsilon_n^2 + \Delta_0^2}\right ] \notag \\
 &\times  \left [   \frac{1}{\sqrt{(\epsilon_n+ \omega_l)^2 + \Delta_0^2}} -\frac{1}{\sqrt{\epsilon_n^2 + \Delta_0^2}} \right ]^2 \Bigg \rbrace.
\end{align}
Similarly, the evaluation of diagrams (2e-2h) yields 
\begin{eqnarray}
Q_H^{(2e-h)} (\omega_l) &=& -2 Q_H^{(2a)}(\omega_l). \label{eq:Q2ehmain}
\end{eqnarray}
 
We now turn our attention to the crossed diagrams (2b-2d). Their contribution is
\begin{equation}
{Q_{H}^{(2b-d)}(\omega_l) =\zeta  \frac{e^2 \Delta_0}{(\omega_l \tau)^2} \beta_{\rm OS} h(\omega_l)} \label{eq:Qcrossedfinal}
\end{equation}
with 
\begin{align}
h(\omega_l) &= T \Delta_0 \sum_n \Bigg \lbrace \left [\frac{\epsilon_n + \omega_l}{\sqrt{(\epsilon_n + \omega_l)^2 + \Delta_0^2}}-\frac{\epsilon_n}{\sqrt{\epsilon_n ^2 + \Delta_0^2}}\right ] \notag \\
& \times \left [\frac{1}{\sqrt{(\epsilon_n + \omega_l)^2 + \Delta_0^2}}-\frac{1}{\sqrt{\epsilon_n ^2 + \Delta_0^2}}\right ]^2\Bigg \rbrace.
\end{align}
In Eq.~\eqref{eq:Qcrossedfinal} we introduced a nonuniversal constant
\begin{align}
\beta_{\rm OS} &= -\frac{\pi}{8} \Big \lbrace 2  \int_0^\infty d\rho \;[\partial_{\rho} J_0(\rho)]^3 \;F(\rho) \notag \\
&+ 3 \int_0^\infty d\rho \;\partial_{\rho}^2 J_0(\rho)\; \partial_{\rho} [J_0(\rho)]^2 \;F(\rho) \Big  \rbrace. \label{eq:beta}
\end{align}
The integration variable $\rho = p_F r$ denotes the distance between the two impurities of diagrams (2b)-(2d). For the general dispersion relation we expect $\beta_{\rm OS} \sim 1$, while for the specific case of a parabolic band we find $\beta_{\rm OS} = 1/8$. Also, note that the integral~\eqref{eq:beta} is determined by lengthscales $r \sim p_F^{-1}$, i.e. by length scales much smaller then the coherence length.

We conclude this section with a comment on the role of the particle-hole (PH) transformation, i.e. of interchanging electronic creation and annihilation operators. As explained in Ref.~[\onlinecite{LutchynYakovenko2009}], for our model this transformation is equivalent to mapping $\xi_{\v p} \rightarrow - \xi_{\v p}$ and $\zeta \rightarrow - \zeta$. It was also shown there, that PH symmetry in the normal phase, i.e. $\xi_{\v p} = -\xi_{\v p}$, combined with the fact $\sigma_{xy} \propto \zeta$ which follows from the time reversal operation, implies $\sigma_{xy} \equiv 0$ for our model with Gaussian disorder. Furthermore, a generic DR, treated in the linearized approximation, $\xi_{\v p} \simeq v_F (p - p_F)$, is PH symmetric upon a redefintion of momenta. By consequence, since the contributions presented in Eqs.~\eqref{eq:Q2amain} and \eqref{eq:Q2ehmain} stem from the Fermi surface, they will vanish whenever $\nu(\xi) = \nu(-\xi)$. Similarly, if the system has PH invariance, $(\xi\mathcal G)(\epsilon, \v r) =- (\xi\mathcal G)(\epsilon, \v r) = 0$ and thus $\beta_{\rm OS}$ and $Q_{H}^{(2b-d)}$, Eqs.~\eqref{eq:beta} and \eqref{eq:Qcrossedfinal}, will be zero in accordance with the general arguments exposed in this section.

\textbf{\textit{Results.}}
Evaluating Matsubara sums in expressions for $h(\omega_l)$ and $k(\omega_l)$ followed by an analytical continuation $i \omega_l \rightarrow \Omega^+ = \Omega + i 0$ one finds
\begin{subequations}
\begin{eqnarray}
k(\omega_l) &\rightarrow& i K (\Omega^+/\Delta_0)\\ \label{eq:analcontK}
h(\omega_l) &\rightarrow& i H (\Omega^+/\Delta_0) \label{eq:analcontH}
\end{eqnarray}
with dimensionless functions 
\begin{widetext}
\begin{align}
K(z) &=  12  z \int_1^\infty \frac{dx}{2\pi} \tanh \left (\frac{\Delta_0 x}{2T}\right ) \frac{1}{\sqrt{x^2 - 1}\left [ 4 x^2-z^2  \right ]} - i\frac{1}{2} \tanh\left (\frac{\Delta_0}{2 T}\right ) \left [\sqrt{\frac{z}{z-2}} + \sqrt{ \frac{z}{z+2 }}\right] \notag \\
& + 2 \int_1^\infty \frac{dx }{2\pi} \tanh\left (\frac{x \Delta_0}{2T}\right ) \sqrt{x^2 -1} \left [\frac{1}{(2x + z)[1 - (x+ z)^2]}-\frac{1}{(2x - z)[1 - (x- z)^2]}\right] , \\
H(z) &= \int_1^\infty dx \frac{\tanh\left (\frac{x \Delta_0}{2T}\right )}{\pi\sqrt{x^2-1}} \left [  \frac{3x + 2z}{1-(x+z)^2}- \frac{3x - 2z}{1-(x-z)^2}\right ]  - \frac{\tanh\left (\frac{\Delta_0}{2 T}\right )}{2} \left [\frac{z - 3}{\sqrt{2z - z^2}} + \frac{z + 3}{\sqrt{-2z - z^2}}\right ].
\end{align}
\end{widetext}
\end{subequations}

In this notation, the contribution of noncrossing diagrams (2a), (2e-2h) to the Hall response is 
\begin{subequations}
\begin{equation}
\sigma_{H}^{(2a,e-h)}  =- \zeta \frac{e^2}{\hbar} 
\frac{\beta_{\rm FS}}{8}
\frac{\Delta_0}{ \Omega^3 \tau^2}  
K (\Omega^+/\Delta_0). \label{eq:sigmaxynoncrossedfinal}
\end{equation}
The contribution of crossed diagrams (2b-2c) to the Hall response, i.e. the major result of this work, is
\begin{equation}
\sigma_{H}^{(2b-d)}  =\zeta \frac{e^2}{\hbar} \beta_{\rm OS} \frac{\Delta_0}{\Omega^3 \tau^2}  H (\Omega^+/\Delta_0) . \label{eq:sigmaxycrossedfinal}
\end{equation}
\label{eq:FinalResult}
\end{subequations}
The total Hall response is the sum of Eqs.~\eqref{eq:sigmaxynoncrossedfinal} and \eqref{eq:sigmaxycrossedfinal}. While above two contributions have very different functional dependence on temperature and on the ac frequency their asymptotic behavior is close. Indeed, for $T = 0$ the limiting cases of functions $H(z)$ and $K(z)$ with $z = \Omega/\Delta_0 + i0$ are
\begin{subequations}
\begin{align}
K(z) &= \begin{cases} \frac{ z^3}{15 \pi} &  \Omega \ll \Delta_0, \\ - i - \frac{6 \ln(z) }{\pi z} & \Omega \gg \Delta_0, \end{cases} \\
H(z) &= \begin{cases} \frac{8 z^3}{105 \pi} &  \Omega \ll \Delta_0, \\ - i - \frac{4  \ln(z) }{\pi z} & \Omega \gg \Delta_0. \end{cases} 
\end{align}
\end{subequations}

\textbf{\textit{Discussion.}}
First, we would like to dwell on the physical meaning of the diagrammatic calculation. In the inset of Fig.~\ref{fig:HallPlot}, the quantum mechanical probability for connecting source and drain, $p =  \sum_{i,j} A_i A_j^*$, is depicted. Amplitudes $A_i$ (their complex conjugate $A_i^*$) are represented in red (blue). Also notice the ``anomalous'' propagation with two opposite arrows on a single line. It represents reflection off the condensate and, as a consequence of averaging over the Fermi surface, occurs in the vicinity of the current vertex.\cite{footnote1} The proportionality $\sigma_H \propto \zeta$ immediately follows. 

Concerning the diffractive, crossed diagrams, Eq.~\eqref{eq:sigmaxycrossedfinal}, recall that those involve a prefactor $\beta_{\rm OS}$ which is determined by the function $F(p_F r)$ and thus stems from virtual (off-shell) processes. As a consequence, by means of Heisenberg's incertainty principle, $\sigma_H^{(2b-d)}$ is determined by impurities residing about one Fermi wavelength $\lambda_F$ from each other, i.e. from impurity complexes represented by yellow areas in Fig.~\ref{fig:HallPlot}. Interestingly, those impurity complexes act similarly to strong impurities. Indeed, diagrams (2b-2d) have the same functional form, Eq.~\eqref{eq:sigmaxycrossedfinal}, as the ``Mercedes star" diagram\cite{Goryo2008, LutchynYakovenko2009} $(1_3)$ in Fig.~\ref{fig:PertDiagsSecond}, which involves the third moment in the distribution of $V(\v r)$. However, in contrast to diagram $(1_3)$, we repeat that $F(p_F r)$ and thus $\sigma_H^{(2b-d)}$ vanish for a strictly PH symmetric model, in accordance with general arguments\cite{LutchynYakovenko2009} reviewed above. On the basis of these considerations, the relative importance of the diffractive contribution, Eq.~\eqref{eq:sigmaxycrossedfinal}, as compared to the previously known result, Eq.~\eqref{eq:sigmaxynoncrossedfinal}, is apparent:
\begin{equation}
{\sigma_{H}^{(2b-d)}}/{\sigma_{H}^{(2a,e-h)}} \sim \beta_{\rm OS}/\beta_{\rm FS}. \label{eq:comparison}
\end{equation}
In a model for which the DOS is nearly constant $\beta_{\rm FS}  \ll 1$ and consequently the diffractive contribution can be parametrically enhanced as compared to other impurity-induced processes computed within ladder approximation. In particular, in the case of a parabolic band, the Hall response is finite, see Fig.~\ref{fig:HallPlot}, as compared to the vanishing result that one obtains from noncrossing diagrams.\cite{LutchynYakovenko2009}

\textbf{\textit{Summary.} } 
The microscopic origin and quantitative understanding of the Kerr effect in TRS-broken state of unconventional superconductors remains as a topic of ongoing debate and active research. Existing calculations in various models and initial assumptions yield very different results concerning the functional dependence of anomalous ac Hall conductance on essential parameters such as frequency, electronic mean free path, and temperature. In particular, in the experimentally relevant frequency range, $\Omega\gg\Delta_0$, two-band model calculations in the clean limit result in the quadratic decay of $\sigma_H$ with inverse frequency $\sigma_H\propto1/\Omega^2$. \cite{TaylorKallin2012} In contrast, impurity based calculations performed either in the model of non-Gaussian disorder or without particle-hole symmetry predict $\sigma_H\propto1/\Omega^3$ scaling from skew-scattering, see Eq.~\eqref{eq:FinalResult} and Refs.~[\onlinecite{Goryo2008, LutchynYakovenko2009}]. However, each of these two extrinsic mechanisms implies different dependence on impurity concentration as $n_{\rm imp}$ and $n_{\rm imp}^2$ respectively. The most recent full $T$-matrix analysis \cite{LiAndreevSpivak2015} uncovered that skew scattering of low-energy quasiparticle on strong impurities results in anomalous Hall response being linearly proportional to $\tau \propto n_{\rm imp}^{-1}$ and falling off linearly with inverse frequency. The kinetic equation approach of Ref. [\onlinecite{LiAndreevSpivak2015}] is however limited to low frequencies, $\Omega<\Delta_0$. Furthermore, since quasiparticle density decreases exponentially fast with temperature, this mechanisms dominates Kerr rotation only in the immediate vicinity of the critical temperature $T_c-T\ll T_c$.   

Apart from its purpose in the context of $p$-wave superconductors, our work can be seen as a proof of principle for the importance of diffractive skew scattering (crossed diagrams) beyond the context of the dc AHE. In particular, our study of the ac AHE evokes a similar investigation for time reversal symmetry breaking superconductors with other unconventional order parameter symmetries. This is also motivated by recent Kerr measurements in high-$T_c$ cuprates YBa$_2$Cu$_3$O$_{6+x}$ \cite{Kapitulnik-YBCO} and La$_{1.875}$Ba$_{0.125}$CuO$_4$, \cite{Kapitulnik-LBCO} and heavy fermion superconductors UPt$_3$ \cite{Kapitulnik-UPt3} and URu$_2$Si$_2$ \cite{Kapitulnik-URuSi}, that already triggered new theories. \cite{Chubukov} We will devote a separate publication to this topic. More generally, it can be expected that diffractive skew scattering plays a substantial role for a plethora of other anomalous physical observables (e.g. the spin Hall effect~\cite{MilletariFerreira2016}) and thus constitutes a focus for future research.

\textbf{\textit{Acknowledgements.}}
We thank A. Andreev, I. Dmitriev, P. Ioselevich, M. Khodas, R. Lutchyn, P. Nagornykh, P. Ostrovsky, D. Pesin, J. Sauls and V. Yakovenko for useful discussions, and acknowledge hospitality by the Department of Physics at University of Michigan (E.J.K.). This work was financially supported in part by NSF Grants No. DMR-1606517 and ECCS-1560732. Support for this research at the University of Wisconsin-Madison was provided by the Office of the Vice Chancellor for Research and Graduate Education with funding from the Wisconsin Alumni Research Foundation.

%

\clearpage
\setcounter{equation}{0}
\setcounter{figure}{0}
\setcounter{table}{0}
\setcounter{page}{1}
\makeatletter
\renewcommand{\theequation}{S\arabic{equation}}
\renewcommand{\thefigure}{S\arabic{figure}}
\renewcommand{\bibnumfmt}[1]{[S#1]}
\renewcommand{\citenumfont}[1]{S#1}

\onecolumngrid

\begin{center}
\textbf{Supplemental Materials to}

\textbf{\large ``Kerr effect from diffractive skew scattering in chiral $p_x \pm i p_y$ superconductors''} \\

Elio J. K\"onig and Alex Levchenko
\end{center}

\section{Derivation of ac Hall conductivity}

\subsection{Density of states}
It was pointed out in [Lutchyn \textit{et al.}, Phys.~Rev.~B 80, 104508 (2009)] that a finite Hall response for a model with Gaussian disorder requires particle-hole asymmetry in the normal state DOS. We remind the reader, that for a quadratic DR, the DOS is
\begin{subequations}
\begin{equation}
\nu(\xi) = \underbrace{\frac{m}{2\pi}}_{=: \nu_0} \theta( \xi+ \mathcal E_F),
\end{equation}
and particle hole symmetry is thus broken at the scale $\mathcal E_F$. For a generic, isotropic $\xi_{\v p}$, the DOS close to the Fermi surface is
\begin{equation}
\nu(\xi) \simeq \underbrace{\frac{p_F}{2\pi v_F}}_{=:\nu_0} \left [1+ \underbrace{\left (1- \frac{p_F}{m_F v_F}\right )\frac{1}{2 \mathcal E_F} }_{=:\nu_0'/\nu_0} \xi\right ].
\end{equation}
\end{subequations}
In the case of generic dispersion relation, we define $\mathcal E_F := v_F p_F/2$. The dispersion relation close to the Fermi surface is $ \xi_p \simeq v_F (p - p_F) + (p - p_F)^2/2m_F$ and we assume an electron like Fermi surface $v_F = \partial_p \xi_p \vert_{p = p_F} >0$.

\subsection{Real space Green's functions}
\subsubsection{Presentation of result}

We will need the real space Green's function $\mathcal G(\epsilon, \v r)$ to zeroth order in $\alpha$. The limit $\alpha \rightarrow 0$ is taken before any other limit, and thus formally $\alpha \ll p_F r$ is assumed. It is determined by the Fermi surface and thus, for any dispersion relation,
\begin{equation}
\mathcal G(\epsilon,\v r)= \frac{\nu_0 \pi}{\sqrt{\epsilon^2 + \Delta^2}}  \lbrace J_0(p_F r) + \mathcal O (\alpha)\rbrace \label{app:eq:realspaceGF}
\end{equation}
Furthermore, we will need $(\xi \mathcal G)(\epsilon, \v r)$ to first order in $\alpha$. It turns out that the zeroth order contribution is determined by the microsopic details at the UV, but the first order contribution is determined by the Fermi surface.
\begin{equation}
(\xi \mathcal G)(\epsilon,\v r) =- {\pi \nu_0} \left \lbrace F(p_F r) + \frac{1}{2}\sqrt{\frac{\epsilon^2 + \Delta_0^2}{\mathcal E_F^2}} \left [\frac{2\mathcal E_F \nu_0'}{\nu_0} J_0(p_Fr) -  p_F r J_1 (p_F r) \right ]\right \rbrace \label{app:eq:realspacexiGF}
\end{equation}
In Eqs.~\eqref{app:eq:realspaceGF} and  \eqref{app:eq:realspacexiGF}, $J_0(p_F r)$ and $J_1(p_F r)$ denote Bessel functions of the first kind. The function $F(p_F r)$ represents an off-shell contribution and therefore depends on microscopic details of the model. In the case of a quadratic dispersion we find $F(p_F r) = Y_0(p_F r)$, where $Y_0(p_F r)$ is the zeroth Bessel function of the second kind. For a general DR, we can write $F(p_F r) = \mathbf H_0(p_F r) + \delta F(p_F r)$. The Struve function $ \mathbf H_0(p_F r)$ stems from the vicinity of the Fermi surface while $\delta F(p_F r)$, which vanishes for large distances, stems from other parts of the spectrum. For $p_F r \gg 1$ microscopic details become unimportant, this is reflected in the asymptotic relation $\mathbf H_0(p_F r) \simeq Y_0(p_F r)$.

\subsubsection{Derivation of real space Green's functions}

We first present the derivation of $\mathcal G(\epsilon, \v r)$ for a general dispersion relation. It will be useful to use the notation $\rho = p_F r$, $\eta_\xi = \sqrt{ \frac{\epsilon^2 + \Delta^2 (p(\xi)/p_F)^2}{\mathcal E_F^2}}$ and denote the lower (upper) band edge by $-D_b \mathcal E_F$ ($D_t \mathcal E_F$). We will omit the difference between $\eta_\xi$ and $\eta_0$ which is of order $\mathcal O(\alpha^2)$.

\begin{eqnarray}
\mathcal G(\epsilon, \v r) &\equiv & \int_{\v p} {e^{i \v p \v r}} \mathcal G(\epsilon, \v p) = \frac{1}{\mathcal E_F}\int_{-D_b}^{D_t} dx { \nu(\mathcal E_F x)} \frac{\langle e^{i p (\mathcal E_F x)  r \cos(\theta)} \rangle_\theta}{ x^2 +\eta_0^2 } \notag \\
&\simeq&  \frac{1}{\sqrt{\epsilon^2 + \Delta^2}} \Big [\lim_{\eta_0 \rightarrow 0} \mathfrak I \int_{-D_b}^{D_t} \frac{dx}{x - i \eta_0}  \nu(\mathcal E_F x)J_0(p(\mathcal E_F x) r)  \notag \\
&& + {\eta_0} \lim_{\eta_0 \rightarrow 0}  \mathfrak R \int_{-D_b}^{D_t} dx  (-\partial_x ) \frac{1}{x - i \eta_0}  \nu(\mathcal E_F x)J_0(p(\mathcal E_F x)r)  \notag \\
&& + \frac{\eta^2_0}{2} \lim_{\eta_0 \rightarrow 0}  \mathfrak I \int_{-D_b}^{D_t} dx  (-\partial_x^2 ) \frac{1}{x - i \eta_0}  \nu(\mathcal E_F x)J_0(p(\mathcal E_F x)r) \Big ] 
\end{eqnarray}

Similarly we find
\begin{eqnarray}
(\xi \mathcal G)(\epsilon, \v r) &\equiv & \int_{\v p} {e^{i \v p \v r}} \xi_{\v p} \mathcal G(\epsilon, \v p) = \int_{-D_b}^{D_t} dx x { \nu(\mathcal E_F x)} \frac{\langle e^{i p (\mathcal E_F x)  r \cos(\theta)} \rangle_\theta}{\left (x+ \delta^2\right )^2 +\eta_0^2 } \notag \\
&\simeq&  \lim_{\eta_0 \rightarrow 0} \mathfrak R \int_{-D_b}^{D_t} \frac{dx}{x - i \eta_0}  \nu(\mathcal E_F x)J_0(p(\mathcal E_F x)r)  \notag \\
&& + {\eta_0} \lim_{\eta_0 \rightarrow 0}  \mathfrak I \int_{-D_b}^{D_t} dx  \partial_x  \frac{1}{x - i \eta_0}  \nu(\mathcal E_F x)J_0(p(\mathcal E_F x)r) 
\end{eqnarray}
%

We see that for $\mathcal G(\epsilon, \v r)$ [$(\xi \mathcal G)(\epsilon, \v r)$] even [odd] powers of $\eta_0$ are determined by the residue at the Fermi surface. On the other hand, the odd [even] terms in $\eta_0$ are determined by off-shell contributions and therefore dependent on microscopic details. For example it depends on the details of the cut-off, which is $D_b = 1$ for quadratic DR. For this concrete case, we present the real space Green's function up to $\mathcal O(\alpha^1)$.
 
\begin{equation}
{\mathcal G} (\epsilon, \v r) =\frac{4 \pi \nu_0}{\mathcal E_F} \underbrace{ \int_{-\infty}^\infty\int_{-\infty}^\infty\frac{dx dy}{(2\pi)^2} \frac{e^{i p_F r x}}{(x^2 + y^2 - 1)^2 + \eta_0^2} }_{=:I(p_F r)} \label{eq:realspaceFGqDR}
\end{equation}
 
To obtain the spatial dependence of $I(\rho)$, we use
\begin{equation}
\cos(\rho x) = \cos(\rho) - 2 \sin\left (\rho \frac{x-1}{2}\right )\sin\left (\rho \frac{x+1}{2}\right ),
\end{equation}
and $ I(0) = (\pi - \arctan(\eta_0))/(4 \pi \eta_0)$, so that
\begin{eqnarray}
I(\rho) &=& \cos(\rho) I(0) - \frac{2}{(2\pi)^2} \int_{-\infty}^\infty dx \sin\left (\rho \frac{x-1}{2}\right )\sin\left (\rho \frac{x+1}{2}\right ) \int_{-\infty}^\infty \frac{dy}{(x^2 + y^2 - 1)^2 + \eta_0^2} \notag \\
&=& \cos(\rho) I(0) - \frac{1}{\eta_0 \pi} \int_{0}^1 dx \frac{\sin\left (\rho \frac{x-1}{2}\right )\sin\left (\rho \frac{x+1}{2}\right ) }{  \sqrt{1-x^2}} - \frac{1}{2\pi} \int_{1}^\infty dx \frac{\sin\left (\rho \frac{x-1}{2}\right )\sin\left (\rho \frac{x+1}{2}\right )}{\sqrt{x^2-1}^3} + \mathcal O(\eta_0) \notag \\
&=&\frac{1}{4 \eta_0}J_0(\rho) - \frac{1}{4\pi} \left (\cos(\rho) + 2 \int_{1}^\infty dx \frac{\sin\left (\rho \frac{x-1}{2}\right )\sin\left (\rho \frac{x+1}{2}\right )}{\sqrt{x^2-1}^3} \right )+ \mathcal O(\eta_0) \notag \\
&=&\frac{1}{4 \eta_0}J_0(\rho) + \frac{1}{8} \rho Y_1(\rho) + \mathcal O(\eta_0). 
\end{eqnarray}

For the parabolic dispersion, we can use the specific form of $\xi_{\v p}$ and differentiate Eq.~\eqref{eq:realspaceFGqDR} to obtain $F(p_F r) = Y_0(p_F r)$ in Eq. (6b) of the main text.

\subsection{Evaluation of diagrams}

All calculations in this supplemental material are performed in Matsubara Green's function technique. It should be understood, that $\epsilon =\epsilon_n$ ($\omega = \omega_l$) are fermionic (bosonic) Matsubara frequencies. We introduce the notation $\v a \wedge \v b =  a_x b_y -a_y b_x $ and restrict ourselves to the case $\zeta = 1$. The case $\zeta = -1$ immediately follows, since anomalous Green's functions are always adjacent to the current vertices. Throughout the calculation we employ the following identity
\begin{eqnarray}
G(\epsilon_1, \v p) G(\epsilon_2, \v p) = \frac{i}{\epsilon_1 - \epsilon_2}  \lbrace G(\epsilon_1, \v p) - G(\epsilon_2, \v p) \rbrace. \label{eq:expansionGF}
\end{eqnarray}

\subsubsection{Diagrams in the non-crossing approximation}
We now investigate diagram (2a). 
\begin{eqnarray}
Q_H^{(2a)}  &=& \frac{e^2 T}{2 (2\pi \nu_0)^4 \tau^2} \sum_\epsilon \int_{\v p_1, \v p_2, \v k} {\v p_1 \wedge \v p_2}   \tr\left  [G(\epsilon, \v p_1) G(\epsilon + \omega, \v p_1) \tau_z G (\epsilon + \omega, \v k) \tau_z G (\epsilon + \omega, \v p_2) G(\epsilon, \v p_2) \tau_z G(\epsilon, \v k) \tau_z\right ]\notag\\
&=& - \frac{e^2 T}{2 (2\pi \nu_0)^4 (\omega \tau)^2} \sum_{\epsilon, \epsilon_1, \epsilon_2} \int_{\v p_1, \v p_2, \v k} {\v p_1 \wedge \v p_2}{(\delta_{\epsilon_1, \epsilon + \omega} - \delta_{\epsilon_1, \epsilon})(\delta_{\epsilon_2, \epsilon + \omega} - \delta_{\epsilon_2, \epsilon})} \notag \\
&& \times \tr\left  [G(\epsilon_1, \v p_1)\tau_z G (\epsilon + \omega, \v k) \tau_z G (\epsilon_2, \v p_2)  \tau_z G(\epsilon, \v k) \tau_z\right ]
\end{eqnarray}
We perform the trace over numerators of the Green's function, cf. Eq.~(5), using the antisymmetrization under $(\epsilon_1, \v p_1) \leftrightarrow(\epsilon_2, \v p_2)$.
We obtain
\begin{eqnarray}
Q_H^{(2a)}  &=& \frac{  e^2 T \Delta_0^2}{(2\pi \nu_0)^4 \omega \tau^2 p_F^2} \sum_\epsilon \int_{\v p_1, \v p_2, \v k}  (\v p_1 \wedge \v p_2)^2 \xi_{\v k} \mathcal G(\epsilon + \omega, \v k) \mathcal{G}(\epsilon, \v k)  [\mathcal{G}(\epsilon, \v p_1)-\mathcal{G}(\epsilon + \omega, \v p_1)][\mathcal{G}(\epsilon, \v p_2)-\mathcal{G}(\epsilon + \omega, \v p_2)] \notag \\
&=& - \frac{e^2 T \Delta_0^2}{{2}(2\pi \nu_0)^4 \tau^2 p_F^2} \sum_\epsilon (2\epsilon + \omega) \left \lbrace \int_{\v p} p^2 \mathcal G(\epsilon, \v p)\mathcal G(\epsilon+\omega, \v p) \right \rbrace^2 \left \lbrace (\xi \mathcal G)(\epsilon+\omega, \v r= 0)-(\xi \mathcal G)(\epsilon,  \v r= 0) \right \rbrace. \label{eq:Q2a}
\end{eqnarray}

For the evaluation of the current correlator of diagrams (2e)-(2h) it is useful to investigate the self-energy at Born level, first. It is momentum independent $\Sigma(\epsilon, \v p) = \Sigma(\epsilon)$ and
\begin{eqnarray}
\Sigma(\epsilon) &=& \frac{1}{2\pi \nu_0 \tau} \int_{\v p'} \tau_z G(\epsilon, \v p') \tau_z = - \frac{1}{2\pi \nu_0 \tau} \left [i \epsilon \mathcal G(\epsilon, \v r = 0) + \tau_z (\xi \mathcal G) (\epsilon, \v r = 0)\right ].
\end{eqnarray}
In particular, the self-energy contains a UV divergent contribution. However, $\sigma_H(\Omega)$ is determined by $\Sigma(\epsilon + \omega)  - \Sigma(\epsilon) $, which is finite.
\begin{eqnarray}
Q_H^{(2e-h)}  &=& \frac{e^2 T}{2 (2\pi \nu_0)^3 \tau} \sum_\epsilon \int_{\v p, \v p'} {\v p \wedge \v p'}  \notag \\
&& \times \Big \lbrace \tr \left [G(\epsilon + \omega, \v p) \tau_z G(\epsilon + \omega, \v p') \Sigma(\epsilon + \omega) G(\epsilon + \omega, \v p') G(\epsilon, \v p') \tau_z G(\epsilon, \v p) \right ] \notag \\
&&- \tr \left [G(\epsilon + \omega, \v p') \Sigma(\epsilon + \omega) G(\epsilon + \omega, \v p')  \tau_z G(\epsilon + \omega, \v p)  G(\epsilon, \v p) \tau_z  G(\epsilon, \v p')  \right ] \Big \rbrace \notag \\
&& - [\omega \rightarrow (-\omega)] \notag \\
&=& \frac{e^2 T}{2 (2\pi \nu_0)^3 \tau \omega^2} \sum_\epsilon \int_{\v p, \v p'} {\v p \wedge \v p'} \tr \big \lbrace \tau_z \left [G(\epsilon, \v p) - G(\epsilon + \omega, \v p) \right ] \tau_z \notag \\
&& \times [G(\epsilon, \v p') \Sigma(\epsilon + \omega) G(\epsilon + \omega, \v p') -G(\epsilon + \omega, \v p') \Sigma(\epsilon + \omega) G(\epsilon, \v p') ] \big \rbrace \notag \\
&& - [\omega \rightarrow (-\omega)] .
\end{eqnarray}
We use
\begin{align}
&G(\epsilon, \v p') \Sigma(\epsilon + \omega) G(\epsilon + \omega, \v p') -G(\epsilon + \omega, \v p') \Sigma(\epsilon + \omega) G(\epsilon, \v p') \notag \\
&= - \frac{\omega \Delta_0}{  \pi \nu_0 \tau p_F} (\v p \wedge \boldsymbol \tau) (\xi \mathcal G) (\epsilon + \omega, \v r =0) \mathcal G (\epsilon, \v p')\mathcal G (\epsilon + \omega, \v p'),
\end{align}
take the trace over Nambu space and obtain
\begin{eqnarray}
Q_H^{(2e-h)} &=& \frac{{2}e^2 T \Delta_0^2}{(2\pi \nu_0)^4 \tau^2 \omega p_F^2} \sum_\epsilon (\xi \mathcal G) (\epsilon + \omega, \v r =0) \int_{\v p, \v p'} ({\v p \wedge \v p'})^2  \left [\mathcal G(\epsilon, \v p) - \mathcal G(\epsilon + \omega, \v p) \right ] \mathcal G(\epsilon, \v p') \mathcal G(\epsilon + \omega, \v p')  \notag \\
&& - [\omega \rightarrow (-\omega)] \notag \\
&=& \frac{ {2} e^2 T \Delta_0^2}{(2\pi \nu_0)^4 \tau^2 \omega p_F^2} \sum_\epsilon (\xi \mathcal G) (\epsilon + \omega, \v r =0) \int_{\v p, \v p'} ({\v p \wedge \v p'})^2  \left [\mathcal G(\epsilon, \v p) - \mathcal G(\epsilon + \omega, \v p) \right ] \mathcal G(\epsilon, \v p') \mathcal G(\epsilon + \omega, \v p')  \notag \\
&& - [\omega \rightarrow (-\omega)]  \notag \\
&=&\frac{e^2 T \Delta_0^2}{ (2\pi \nu_0)^4 \tau^2 p_F^2} \sum_\epsilon (2\epsilon + \omega) \left \lbrace \int_{\v p} p^2 \mathcal G(\epsilon, \v p)\mathcal G(\epsilon+\omega, \v p) \right \rbrace^2  \left \lbrace (\xi \mathcal G)(\epsilon+\omega, \v r= 0)-(\xi \mathcal G)(\epsilon,  \v r= 0) \right \rbrace. \label{eq:Q2eh}
\end{eqnarray}

We see, that the functional form of contributions from diagrams (2a), Eq.~\eqref{eq:Q2a}, and (2e)-(2h), Eq.~\eqref{eq:Q2eh}, is the same and that $Q_H^{(2a)} = -Q_H^{(2e-h)}/2$. We use
\begin{eqnarray}
\int_{\v p} p^2 \mathcal G(\epsilon, \v p)\mathcal G(\epsilon+\omega, \v p) &=& \frac{\pi \nu_0 p_F^2}{\omega (2\epsilon + \omega)} \left [ \frac{1}{\sqrt{\epsilon^2 + \Delta_0^2}} -  \frac{1}{\sqrt{(\epsilon+ \omega)^2 + \Delta_0^2}} \right ], \notag \\
(\xi \mathcal G)(\epsilon+\omega, \v r= 0)-(\xi \mathcal G)(\epsilon,  \v r= 0) &=& - \pi \nu_0' \left [\sqrt{(\epsilon + \omega)^2 + \Delta_0^2} - \sqrt{\epsilon^2 + \Delta_0^2}\right ], \notag
\end{eqnarray}
and finally obtain for the current current correlator in the non-crossing approximation 
\begin{equation}
Q_H^{(2a,e-h)} = {-} \frac{e^2 \Delta_0}{2 \omega^2 (2 \tau)^2} \frac{\mathcal E_F \nu_0'}{\nu_0} k(\omega).
\end{equation}

For the case of parabolic dispersion, $\nu_0' = 0$ and thus the contribution of non-crossing diagrams vanishes. This can be checked explicitly, since
\begin{eqnarray}
(\xi \mathcal G)(\epsilon+\omega, \v r= 0)-(\xi \mathcal G)(\epsilon,  \v r= 0) &\equiv & \int_{\v k} \xi_{\v k} [\mathcal G(\epsilon+\omega, \v k) - \mathcal G(\epsilon, \v k)] \notag \\
&=& - \frac{\nu_0}{2} \left [\ln[\mathcal E_F^2 + (\epsilon + \omega)^2 + \Delta_0^2]-\ln[\mathcal E_F^2 + \epsilon^2 + \Delta_0^2]\right ].
\end{eqnarray}
As the summation over $\epsilon$ is dominated by $\epsilon < \omega$, we find that, for the model with parabolic dispersion, the non-crossed diagrams are a factor of $\alpha$ smaller than the crossed diagrams and thus negligible.

This concludes the derivation of Eqs.~(7),(10) of the main text.

\subsubsection{Crossed diagrams}

We now turn our attention to the crossed diagrams (2b)-(2d).
For diagram (2b) we find
\begin{eqnarray}
Q_{H}^X &=& \frac{e^2}{2(2\pi \nu_0)^4 \tau^2} T \sum_{\epsilon} \int_{\lbrace \v p_i \rbrace} (2 \pi)^2 \delta (\v p_1 + \v p_2 - \v p_3 - \v p_4) {\v p_1 \wedge \v p_2} \notag \\
&& \times \tr \left [G(\epsilon, \v p_1) G(\epsilon + \omega, \v p_1)\tau_z G(\epsilon+ \omega, \v p_3) \tau_z G(\epsilon + \omega, \v p_2)G(\epsilon, \v p_2) \tau_z G(\epsilon, \v p_4) \tau_z\right ] \notag \\
&=& - \frac{e^2}{4 (2\pi \nu_0)^4 (\omega \tau)^2} T \sum_{\epsilon, \lbrace \epsilon_i \rbrace}  {(\delta_{\epsilon_1, \epsilon + \omega} - \delta_{\epsilon_1, \epsilon})(\delta_{\epsilon_2, \epsilon + \omega} - \delta_{\epsilon_2, \epsilon})} \left (\delta_{\epsilon_3, \epsilon + \omega}\delta_{\epsilon_4, \epsilon}-\delta_{\epsilon_4, \epsilon + \omega}\delta_{\epsilon_3, \epsilon}\right )\notag \\
&& \int_{\lbrace \v p_i \rbrace}(2 \pi)^2 \delta (\v p_1 + \v p_2 - \v p_3 - \v p_4) \v p_1 \wedge \v p_2 \tr[G(\epsilon_1, \v p_1) \tau_zG(\epsilon_3, \v p_3) \tau_zG(\epsilon_2, \v p_2) \tau_zG(\epsilon_4, \v p_4) \tau_z], \notag \\
\end{eqnarray}
while for the sum of diagrams (2c) and (2d) we obtain
\begin{eqnarray}
Q_{H}^\Psi &=& \frac{e^2}{2(2\pi \nu_0)^4 \tau^2} T \sum_{\epsilon} \int_{\lbrace \v p_i \rbrace} (2 \pi)^2 \delta (\v p_1 - \v p_2 - \v p_3 + \v p_4) {\v p_1 \wedge \v p_2} \notag \\
&& \times \Big \lbrace \tr \left [G(\epsilon, \v p_1) G(\epsilon + \omega, \v p_1)\tau_z G(\epsilon+ \omega, \v p_3) \tau_z G(\epsilon + \omega, \v p_4) \tau_z G(\epsilon + \omega, \v p_2)G(\epsilon, \v p_2) \tau_z\right ] 
\notag \\
&&+ \tr \left [G(\epsilon, \v p_1) G(\epsilon + \omega, \v p_1) \tau_z G(\epsilon + \omega, \v p_2)G(\epsilon, \v p_2)  \tau_z G(\epsilon, \v p_4) \tau_z G(\epsilon, \v p_3)  \tau_z \right ]  \Big \rbrace \notag \\
&=& - \frac{e^2}{2 (2 \pi \nu_0)^4 (\omega \tau)^2} T \sum_{\epsilon, \lbrace \epsilon_i \rbrace}  {(\delta_{\epsilon_1, \epsilon + \omega} - \delta_{\epsilon_1, \epsilon})(\delta_{\epsilon_2, \epsilon + \omega} - \delta_{\epsilon_2, \epsilon})} \left (\delta_{\epsilon_3, \epsilon + \omega}\delta_{\epsilon_4, \epsilon+\omega}-\delta_{\epsilon_3, \epsilon}\delta_{\epsilon_4, \epsilon}\right )\notag \\
&& \int_{\lbrace \v p_i \rbrace}(2 \pi)^2 \delta (\v p_1 - \v p_2 - \v p_3 + \v p_4) \v p_1 \wedge \v p_2 \tr[G(\epsilon_1, \v p_1) \tau_zG(\epsilon_3, \v p_3)\tau_zG(\epsilon_4, \v p_4) \tau_zG(\epsilon_2, \v p_2)  \tau_z]. \notag \\
\end{eqnarray}
We evaluate the traces and find for $X$- and $\Psi-$diagrams
\begin{subequations}
\begin{align}
Q_{H}^X &= \frac{e^2\Delta^2 T}{(\omega \tau)^2}  \sum_{\epsilon, \lbrace \epsilon_i \rbrace}  {(\delta_{\epsilon_1, \epsilon + \omega} - \delta_{\epsilon_1, \epsilon})(\delta_{\epsilon_2, \epsilon + \omega} - \delta_{\epsilon_2, \epsilon})} \left (\delta_{\epsilon_3, \epsilon + \omega}\delta_{\epsilon_4, \epsilon}-\delta_{\epsilon_4, \epsilon + \omega}\delta_{\epsilon_3, \epsilon}\right ) I (\lbrace \epsilon_i \rbrace) ,\\
Q_{H}^\Psi &= \frac{e^2 2 \Delta^2 T}{(\omega \tau)^2}  \sum_{\epsilon, \lbrace \epsilon_i \rbrace}  {(\delta_{\epsilon_1, \epsilon + \omega} - \delta_{\epsilon_1, \epsilon})(\delta_{-\epsilon_2, \epsilon + \omega} - \delta_{-\epsilon_2, \epsilon})} \left (\delta_{\epsilon_3, \epsilon + \omega}\delta_{-\epsilon_4, \epsilon+\omega}-\delta_{-\epsilon_4, \epsilon }\delta_{\epsilon_3, \epsilon}\right ) I (\lbrace \epsilon_i \rbrace).
\end{align}
\end{subequations}
Here, we introduced the integral
\begin{align}
I (\lbrace \epsilon_i \rbrace) &= \frac{1}{(2 \pi \nu_0)^4 p_F^2} \int_{\lbrace \v p_i \rbrace} (2 \pi)^2 \delta (\v p_1 + \v p_2 - \v p_3 - \v p_4) {\v p_1 \wedge \v p_2}  \prod_{i=1}^4 \mathcal G(\epsilon_i, \v p_i) \notag \\
& \times \Big \lbrace \v p_1 \wedge \v p_2 \epsilon_3 \xi_4  -\v p_3 \wedge \v p_4 \epsilon_1 \xi_2  \notag \\
& - \v p_3 \wedge \v p_2 (\epsilon_1 \xi_4+\epsilon_4 \xi_1)  - \v p_1 \wedge \v p_3 (\epsilon_2 \xi_4+\epsilon_4 \xi_2)\Big \rbrace. \notag \\
&= \frac{\pi \beta_1}{8} \left  [\frac{\epsilon_4}{\prod_{i = 1,3,4} \sqrt{\epsilon_i^2 + \Delta_0^2}}+\frac{\epsilon_4}{\prod_{i = 2,3,4} \sqrt{\epsilon_i^2 + \Delta_0^2}}-\frac{\epsilon_3}{\prod_{i = 1,2,3} \sqrt{\epsilon_i^2 + \Delta_0^2}}\right ] \notag \\
&- \frac{\pi \beta_2}{8} \left [\frac{\epsilon_4}{\prod_{i = 1,3,4} \sqrt{\epsilon_i^2 + \Delta_0^2}}+\frac{\epsilon_4}{\prod_{i = 2,3,4} \sqrt{\epsilon_i^2 + \Delta_0^2}}-\frac{\epsilon_1+\epsilon_2}{\prod_{i = 1,2,3} \sqrt{\epsilon_i^2 + \Delta_0^2}}\right  ] \label{eq:CrossedIntegral}
\end{align}
The two constants $\beta_{1,2}$ are defined by the following integrals
\begin{subequations}
\begin{align}
\beta_1 &= \int_0^\infty d\rho  \;\partial_{\rho }^2 J_0(\rho )\; \partial_{\rho } [J_0(\rho )]^2 \;F(\rho ) &\stackrel{\rm quad. DR}{=}& 0  \\
\beta_2 &=- \int_0^\infty d\rho  \;[\partial_{\rho } J_0(\rho )]^3 \;F(\rho ) &\stackrel{\rm quad. DR}{=} &1/2\pi. 
\end{align}
\end{subequations}
Only the last term of the square brackets in Eq.~\eqref{eq:CrossedIntegral} enters the current-current correlator of X- and $\Psi$ diagrams. This is because the previous terms are $\epsilon_1$ or $\epsilon_2$ independent, which in turn follows from $(\xi \mathcal G ) (\epsilon+ \omega, \v r) - (\xi \mathcal G ) (\epsilon, \v r) \sim \alpha$, see also [Lutchyn \textit{et al.}, Phys.~Rev.~B 80, 104508 (2009)], Eq.~(56).

With the help of these expressions, we find the following formula for the current-current correlator
\begin{equation}
{Q_{H}^{X + \Psi} = \frac{e^2 \Delta_0}{(\omega \tau)^2} \beta_{\rm OS} h(\omega)} \label{eq:Qcrossedfinal}
\end{equation}
The constant $\beta_{\rm OS} = {\pi}[2 \beta_2 - 3 \beta_1]/{8}$ is $\beta_{\rm OS} = 1/8$ for the case of quadratic dispersion and generally expected to be of order unity. This concludes the derivation of Eq.~(11) of the main text.


\begin{thebibliography}{34}
\bibitem{footnote1} {In processes (2e-h), the anomalous propagation can also occur in the trajectory connecting the two impurities.}
\end{thebibliography}
\end{document}